\documentclass[apjl,twocolumn]{openjournal}

\makeatletter
\long\def\frontmatter@abstractheading{%
  \centerline{\itshape\footnotesize\@submitted}
  \begingroup
    \centering
    \abstractname
    \par
  \endgroup
  \everypar{\rightskip=0.5in\leftskip=\rightskip}\par
}
\makeatother

\shorttitle{Paths to Stardom}
\shortauthors{Luisi et al}

\journalinfo{The Open Journal of Astrophysics}
\usepackage{amsmath,amsfonts}
\usepackage{hyperref}
\hypersetup{colorlinks=true, citecolor=blue, linkcolor=blue, urlcolor=blue}
\usepackage{xcolor}
\usepackage{orcidlink}

\def\frontmatter@title@above{%
  \vspace*{0pt}%
  \footnotesize
  {\footnotesize\textsc{\@journalinfo}}\par
  {\scriptsize Preprint typeset using \LaTeX\ style openjournal v.\ \openjournaltemplate@ver}\par
}%
\def\frontmatter@title@below{\vspace*{18in}}%
\makeatother

\begin{document}

\title{Pink Dwarfs and the Paths to Stardom: How Brown Dwarfs Pushed Above the Hydrogen Burning Limit Evolve \vspace{-0.5in}}
\author{Jaime Luisi\,\orcidlink{0009-0000-0251-2892}}
\author{John C. Forbes\,\orcidlink{0000-0002-1975-4449}}
\author{Heather V. Rusk\,\orcidlink{0009-0008-4020-0433}}
\author{Benjamin Gullick\,\orcidlink{0009-0003-6467-5058}}
\affiliation{School of Physical and Chemical Sciences--Te Kura Mat\=u, University of Canterbury, Christchurch 8140, New Zealand}


\begin{abstract} 
Brown dwarfs that gain mass through binary interactions may be pushed above the boundary that divides brown dwarfs from low-mass stars: the hydrogen burning limit (HBL). Some of these objects will make their way to the main sequence and may eventually be indistinguishable from ordinary low-mass stars, while others will remain brown dwarf-like, unable to burn hydrogen at a high enough rate to power their surface luminosity. We study the evolution of both types of object to provide a taxonomy and testable observational predictions for these objects depending on their evolutionary path. Using MESA simulations, we find that a subset of the objects that will eventually become stars experience an extended luminosity plateau, where their surface luminosity remains nearly constant on 100 Myr - Gyr timescales. We find that the plateau timescale is set by the amount of energy required to re-heat the cores of these objects to a level sufficient to sustain convection. The timescales required for the cores of these objects to ``unfreeze'' and arrive at the main sequence is long enough that surveys may be able to find objects in this evolutionary stage. These objects, along with those that never reach the main sequence, occupy a unique space in a mass-luminosity diagram, and would provide a unique constraint on binary mass transfer physics.
\end{abstract}

\section{Introduction} \label{introduction}
Brown dwarfs are popularly known as ‘failed stars’. They have too little mass to stably burn hydrogen in their cores, so they exist below the hydrogen burning limit (HBL). The existence of brown dwarfs was first predicted theoretically by S. Kumar in the 1960s \citep{Kumar1, Kumar2, Kumar3, Kumar4, Kumar5}, and they were definitively discovered over 30 years later \citep{Nakajima, Oppenheimer}. Brown dwarfs have served as important test cases for the physics of stellar and planetary atmospheres given the complexities of cloud formation and chemistry \citep[e.g.][]{1997Burrows,1997Allard,2001Allard,2011Allard,2012Allard,Morley2012,Marley2021}.

\begin{figure*}
\centering
\includegraphics[width=0.7\textwidth]{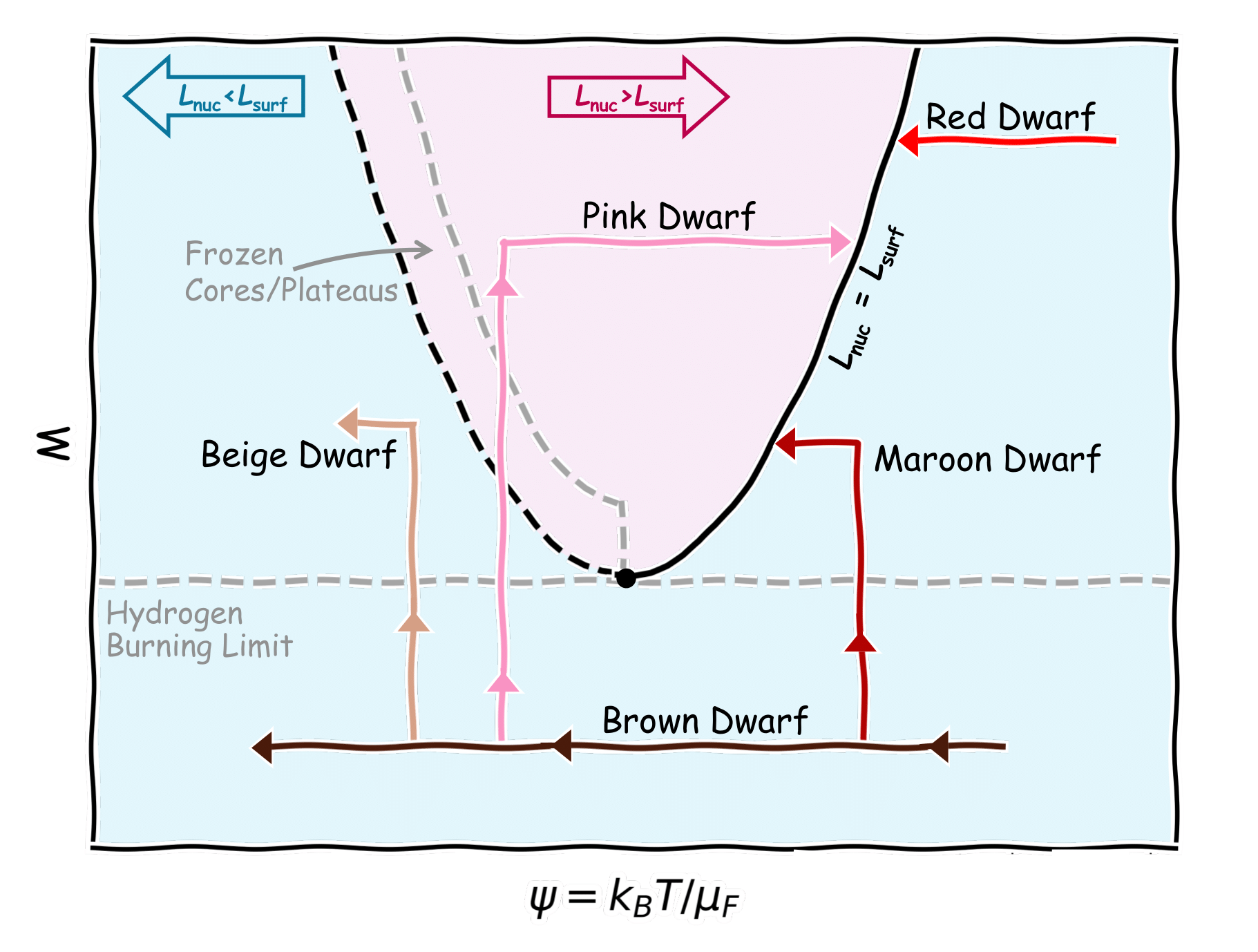}
\caption{A schematic diagram. The coloured arrows represent the paths the different stellar and substellar objects take as they evolve on a plot of mass against $\psi$, the thermal energy per particle in units of the Fermi energy in the star's core. Lower values of $\psi$ indicate colder and more degenerate cores. The separately coloured regions and their corresponding outline arrows indicate what direction an object would move. In the pink region within the parabola, objects have a nuclear luminosity, $L_{\mathrm{nuc}}$, greater than their surface luminosity, $L_{\mathrm{surf}}$ and they evolve to the right as their cores heat up. In the blue region, $L_{\mathrm{surf}} > L_{\mathrm{nuc}}$, so objects evolve to the left as they radiate more energy than they generate. The black parabola is where $L_{\mathrm{nuc}}=L_{\mathrm{surf}}$. The solid right-hand side of the parabola represents the main sequence; it is stable to perturbations in $\psi$. The dashed left-hand side is unstable. The lowest mass on the parabola is the hydrogen burning limit, M$_\mathrm{HBL}$. The arrows for the red dwarf and brown dwarf are horizontal as these objects do not change their mass throughout their lives without an outside influence. Maroon, pink, and beige dwarfs begin as brown dwarfs and evolve leftwards until an episode of mass accretion results in them splitting off from the brown dwarf arrow vertically, at least in the case of fast accretion. When accretion has stopped, they then evolve horizontally to the left or right. Maroon dwarfs, like red dwarfs, evolve left and settle on the main sequence. Beige dwarfs end up above the hydrogen burning limit, but outside of the parabola, and so evolve left like a brown dwarf, never reaching the main sequence. A pink dwarf is located inside the parabola when mass has stopped accreting and then evolves to the right to settle on the main sequence. The `frozen cores/plateaus' region in the parabola is the area in which a subset of our pink dwarf models have a frozen core or luminosity plateau (see Section \ref{sec:results}).}
\label{fig:schematic}
\end{figure*}

Brown dwarfs are typically distinguished from low-mass stars by their mass. Objects below the HBL mass $M_\mathrm{HBL} \sim 0.075\ M_\odot$ at Solar metallicity\footnote{As metallicity increases, $M_\mathrm{HBL}$ decreases \citep{Kumar4,Saumon1994}} \citep{2023Chabrier} may be able to burn hydrogen in their cores, but not enough to supply the luminosity they radiate at their surface ($L_\mathrm{surf}$). Objects with masses exceeding $M_\mathrm{HBL}$ however, will reach the main sequence, at which point their surface luminosity will be supplied solely by the luminosity of hydrogen burning ($L_\mathrm{nuc}$). Just above the HBL stars can remain in this stable hydrogen-burning state for $\sim 10^{13}$ years, whereas just below $M_\mathrm{HBL}$, objects steadily cool, contract, and become more degenerate \citep{1997Burrows,JohnLoeb}.

While this mass-based distinction between brown dwarfs and stars holds broadly, binary interactions can complicate the picture. In particular, if a brown dwarf gains enough mass that it exceeds the HBL some time after its formation, it may take one of several paths depending on the degeneracy of its core at the time. If the core is sufficiently degenerate, \citet{JohnLoeb} showed that the object would remain essentially brown dwarf-like, cooling and contracting, despite exceeding the HBL. They called these objects overmassive brown dwarfs.

\citet{JohnLoeb} proposed gravitational wave driven mass transfer in binary brown dwarf systems via Roche lobe overflow as the most plausible mechanism by which overmassive brown dwarfs could be formed astrophysically. This conclusion was based on the requirement that the mass transfer needed to happen gigayears after the formation of the brown dwarf to allow its core to become sufficiently degenerate. \citet{DorsaJohn} showed, however, that this mechanism requires some fine-tuning because the brown dwarfs need to be extremely close together to merge in a Hubble time, in fact closer than the sizes of the brown dwarfs early in their lives. While this formation channel is still a possibility with dynamical capture or hardening of a tight binary, the intrinsic rarity of brown dwarf-brown dwarf binaries \citep{close2003, burgasser2006, gelino2011, fontanive2018,fontanive2023} and the additional dynamical processing necessary, suggests that this channel should be rare. \citet{DorsaJohn} instead considered the case of mass transfer from the slow winds of an asymptotic giant branch star being accreted by a brown dwarf companion, which can plausibly add enough mass to a brown dwarf to push it over the HBL gigayears after the formation of the brown dwarf (the main sequence lifetime of the donor star). This mechanism would result in an overmassive brown dwarf in a binary with the donor star's white dwarf remnant.

Overmassive brown dwarfs had also been discussed in the 1990s as a dark matter candidate, with \citet{Hansen} coining the term ``beige dwarfs'' for such objects. \citet{Salpeter}, \citet{Hansen}, and \citet{Lyden-Bell} each considered similar possibilities, though through a somewhat different mechanism than \citet{JohnLoeb} and \citet{DorsaJohn}. In the former, beige dwarfs were produced through thermal instability from low-mass seeds and therefore remained cold provided the accretion rates remained small enough \citep{Lenzuni1992}, while in the latter case and in this work, we consider brown dwarfs formed via ordinary star formation. In the latter case, the brown dwarfs need to spend substantial time cooling to get rid of the thermal energy obtained in their formation. Interest in beige dwarfs, regardless of their particular formation mechanism, waned when they were ruled out as the dominant component of dark matter \citep{Alcock}. They are still interesting in their own right however as objects that are uniquely sensitive to the physics of binary mass transfer  \citet{DorsaJohn}. In general, binary systems that have undergone mass transfer retain little information about the state of the system prior to the mass transfer, but in the case of beige dwarfs, the beige dwarf must have had $M<M_\mathrm{HBL}$ prior to the mass transfer. 

To become beige dwarfs, brown dwarfs must wait multiple gigayears for their cores to become sufficiently degenerate such that, at the time of the mass transfer, their mass can be pushed above $M_\mathrm{HBL}$ without kickstarting a much higher nuclear fusion rate in the process. Presumably if the mass transfer happens earlier, the object will not become a beige dwarf, but rather a star. This is the process we explore in this work. We aim to clarify where this transition takes place, and whether there is any observable difference between ordinary low-mass stars and ones that have undergone this process of mass transfer.

Our starting point for understanding this evolution is Figure \ref{fig:schematic}, a diagram of the mass of an object against $\psi = k_B T/\mu_F$, its temperature in units of the Fermi energy, which is a measure of how degenerate the core of the star is (lower values of $\psi$ are more degenerate). In this diagram, the locus of points where the nuclear luminosity is equal to the surface luminosity is roughly parabolic (hereafter ``the parabola''). The minimum of this set of points defines the hydrogen burning limit, the lowest mass at which an object can burn hydrogen at a rate that supplies the surface luminosity. The parabola-like shape can be derived semi-analytically \citep{auddy2016} by assuming that the object has a nearly constant entropy, which sets the structure of the star and therefore both the surface and nuclear luminosities at a given mass and entropy (closely related to the parameter $\psi$). While the exact shape and location of the parabola is sensitive to the treatment of the atmosphere, numerical models \citep{JohnLoeb} are ``aware'' of the parabola, with low-mass stars evolving from the right-hand side of the diagram and settling onto the parabola, corresponding to the zero-age main sequence, while brown dwarfs continuously become more degenerate (decreasing $\psi$) as they pass wholly below the parabola.

Outside of (below) the parabola, the nuclear luminosity is less than the surface luminosity, so the object loses energy and cools. Within the parabola, the opposite is true. This suggests that the left-hand side of the parabola represents an unstable equilibrium, where any perturbation in $\psi$ takes the object away from $L_\mathrm{nuc}=L_\mathrm{surf}$. Objects more degenerate than the $L_\mathrm{nuc} = L_\mathrm{surf}$ line will therefore just continue cooling forever like brown dwarfs, and indeed these objects can be formed in numerical calculations with stellar evolution codes \citep{JohnLoeb, DorsaJohn}.

Despite its clear simplifications, i.e. the assumption that all objects are fully convective and therefore have essentially constant entropy, this diagram has been remarkably successful at explaining these numerical results, namely the existence of beige dwarfs, the hydrogen burning limit, and the shape of the low-mass tail end of the main sequence in this space. We therefore expect that it will be a good guide to understanding what happens when an object ends up, via mass transfer, within the parabola or to its right. Objects within the parabola should heat up and eventually find their way to the main sequence, while objects to the right should likewise end up on the main sequence, following the same path as ordinary low-mass pre-main-sequence stars. We call the former ``pink dwarfs'' and the latter ``maroon dwarfs.''

In Section \ref{sec:method} we describe the setup of a series of numerical simulations to simulate brown dwarfs to which we add mass. In Section \ref{sec:results} we show what happens to these objects and why. We discuss the prospects for identifying beige and pink dwarfs observationally in section \ref{sec:discussion}, and conclude in section \ref{sec:conclusion}.

\section{Simulation Setup} \label{sec:method}

To model the evolution and mass accretion of brown dwarfs and low-mass stars, we have used the open-source stellar evolution code MESA, Modules for Experiments in Stellar Astrophysics \citep[MESA][]{Paxton2011, Paxton2013, Paxton2015, Paxton2018, Paxton2019, MESA}.
Stellar models have found difficulty in modelling brown dwarfs due to the complicated cloud physics and molecular chemistry in their atmospheres \citep[e.g.][]{2012Allard}. With this in mind, when using MESA for the modelling of brown dwarfs we have set the boundary condition deeper into the atmosphere, using the tau 100 tables following \cite{Choi}. Full coupling of interior models with sophisticated atmospheric models is possible and makes some difference in retrieval estimates of brown dwarf properties \citep{gerasimov2024jwst,gerasimov2024exploring}, but we ignore this complication here.

For each potential overmassive dwarf/low mass star, we ran and then concatenated three separate runs of MESA\footnote{Version r24.03.1}, which we refer to as Run 1, Run 2, and Run 3 respectively. Run 1 evolves a brown dwarf with an initial mass below the HBL. Run 2 loads the model from Run 1, and adds mass to it over a period of time. Run 3 loads the model created at the end of Run 2 and, without adding any more mass, lets the object evolve until $10^{11}$ years (with respect to the time Run 1 began). This long time limit allows us to unambiguously classify the state of the object as a star (as opposed to something brown dwarf-like) based on the relative values of the surface and nuclear luminosities.

\begin{figure}
    \centering
    \includegraphics[width=\linewidth]{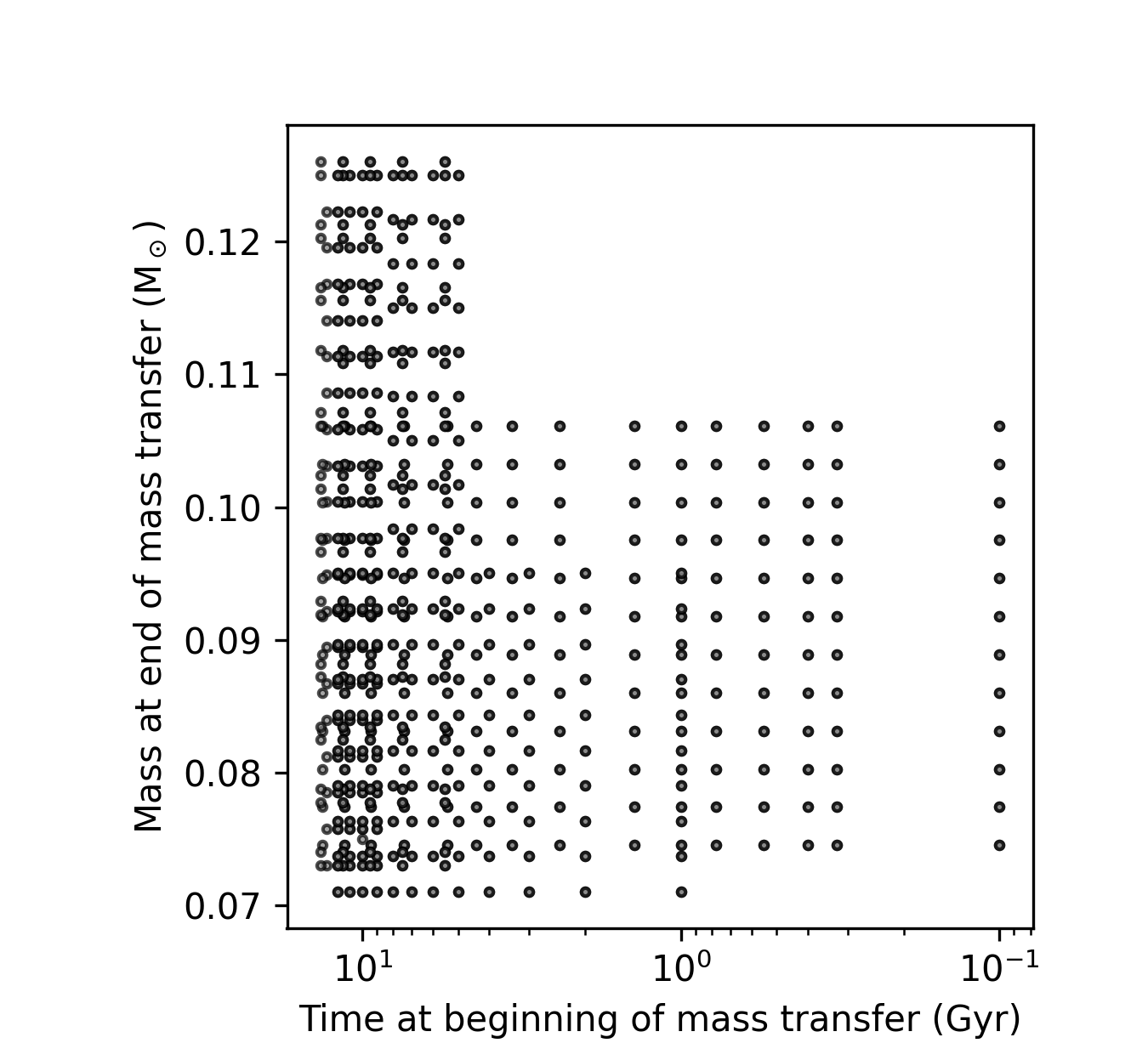}
    \caption{The MESA runs. In each run, the first key parameter is the time at which a brown dwarf begins a mass accretion phase, measured relative to its formation (the x-axis, which is reversed so that the brown dwarf moves leftward just like in Figure \ref{fig:schematic}). The second parameter is the final mass of the brown dwarf after its accretion phase, which in all cases begins as a $0.065\ M_\odot$ ordinary brown dwarf. Note that for each point in this figure we run a case where the accretion is fast, taking $10^7$ years, and a case where the accretion is slow, taking $10^9$ years.}
    \label{fig:runs}
\end{figure}

Our runs begin as a straightforward modification of the \texttt{make\_brown\_dwarf} case from the MESA test suite.
Run 1 is a brown dwarf of $0.065 M_\odot$ that we run for varying lengths of time, which allows its core to reach differing levels of degeneracy. In Run 2, mass is added at a constant rate. We parameterise the rate as $\dot{M} = \Delta M/\Delta t$, with $\Delta M$ varied systematically to yield various final masses, and $\Delta t$ set to either $10^7$ years (``fast accretion'') or $10^9$ years (``slow accretion''). The former is more typical for accretion from an approximately Solar mass star undergoing giant-phase evolution \citep{DorsaJohn}, while the latter more closely resembles the binary brown dwarf scenario proposed by \citet{JohnLoeb}. In Run 3, the accretion rate is set back to zero and the star is allowed to evolve until it is clear what sort of steady state it has reached.

As a point of comparison, we also run a series of ``vanilla'' cases, where we evolve objects with no accretion phase so that we can identify the main sequence and isochrones for ordinary objects in this mass range. These runs are identical to ``Run 1'' above, but with no specified end time, and with a varying initial mass.

In Figure \ref{fig:runs} we show the set of runs where we add mass to a $0.065\ M_\odot$ brown dwarf. This set was constructed with a series of regular grids to provide more details in interesting parts of the parameter space as the results were analysed. The time on the x-axis is measured relative to the formation of the brown dwarf, and has a 1:1 correspondence with the value of $\psi$ in the schematic diagram presented in the introduction. Runs are also parameterised by how much mass is added in Run 2, yielding the final mass shown on the y-axis.

\section{Results}
\label{sec:results}

\begin{figure*}
\centering 
\includegraphics[scale=0.7]{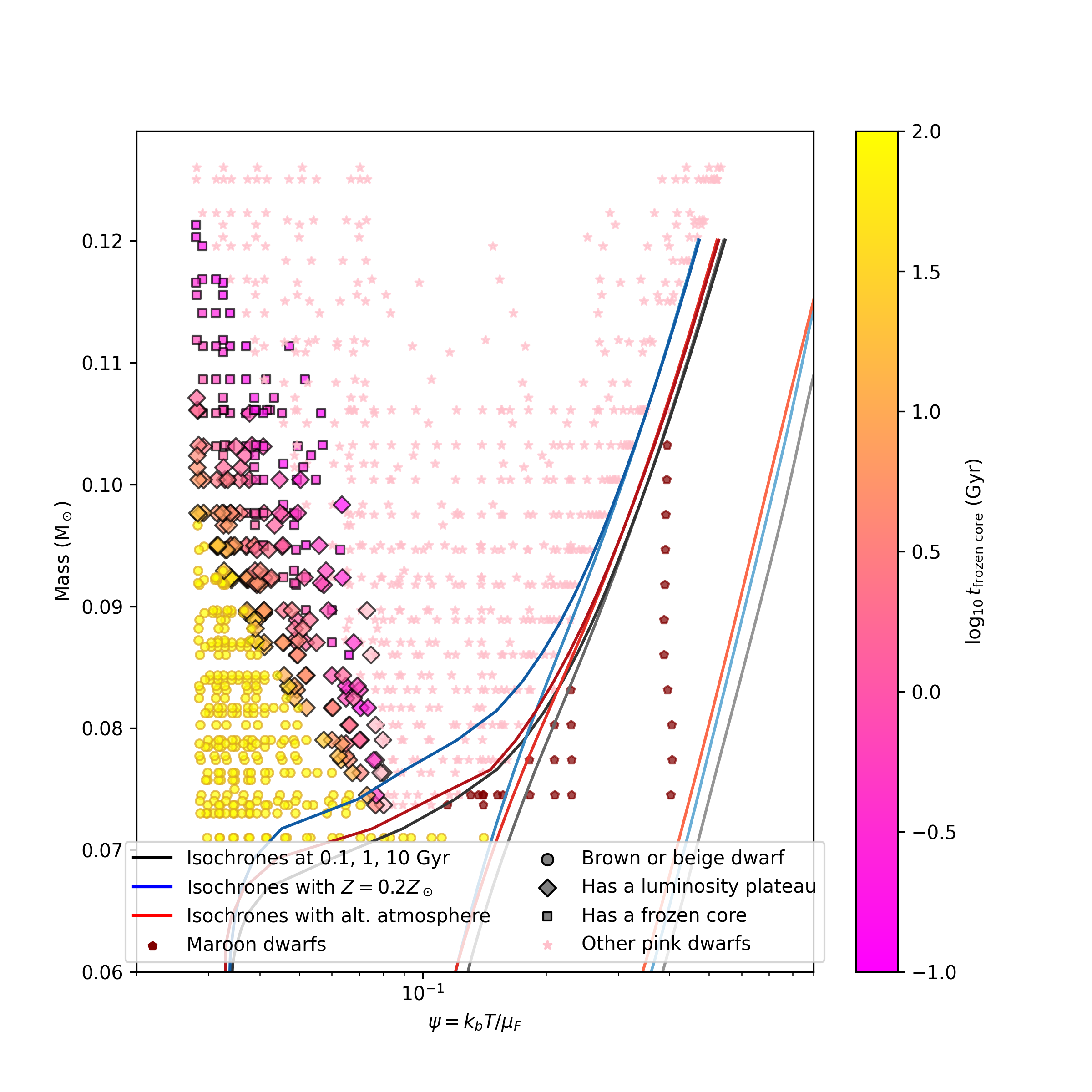}
\caption{MESA models in the $M-\psi$ diagram.
MESA models at the end of their accretion phase are plotted in the same space as the schematic diagram (Figure \ref{fig:schematic}): mass versus the core temperature in units of the Fermi energy, which is 1/\texttt{center\_degeneracy} in terms of MESA outputs. Models are assigned colours and symbols according to their final fates. Models that never reach the main sequence are yellow circles (beige dwarfs). Models that approach the main sequence from the right are maroon pentagons (maroon dwarfs). Pink dwarfs that have a frozen core (see text) are squares; pink dwarfs that have luminosity plateaus (see text) are diamonds, and  pink dwarfs that have neither are pink stars. We also show several isochrones for the ordinary evolution of low-mass stars and brown dwarfs at 0.1, 1, and 10 Gyr from right to left, which show the presence of the main sequence and the continuous cooling of ordinary brown dwarfs. Isochrones for several versions of these models are shown: the red lines have altered atmospheric boundary conditions, and the blue lines are models with lower metallicities.}
\label{fig:massentropy}
\end{figure*} 

\begin{figure*}
\centering 
\includegraphics[scale=0.8]{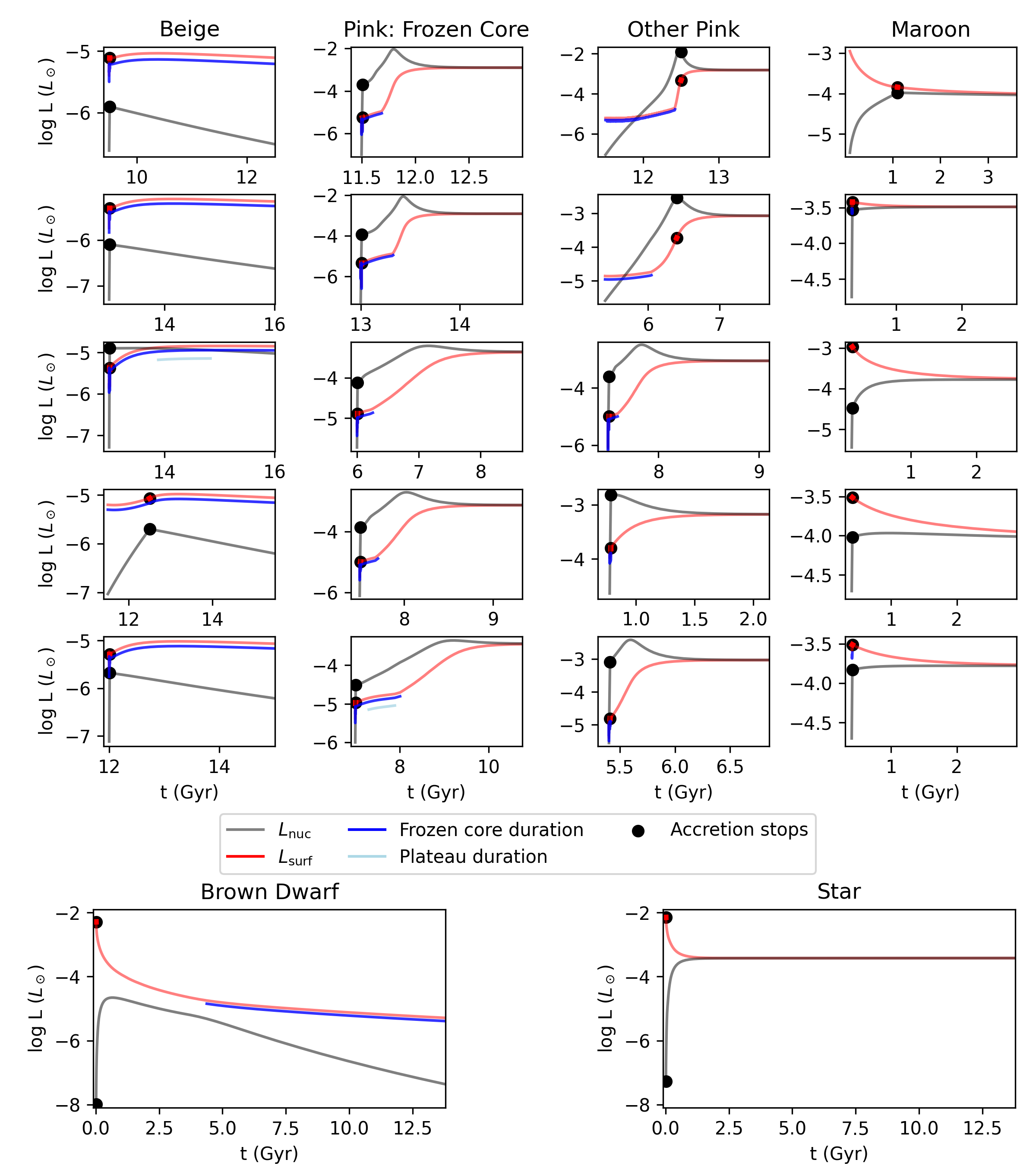}
\caption{Luminosity over time of an example set of objects. The top five rows show examples of brown dwarfs to which mass was added, while the bottom two panels show a brown dwarf and a star with constant masses ($0.065 M_\odot$ and $0.085 M_\odot$ respectively). The luminosity produced by nuclear reactions $L_\mathrm{nuc}$ is shown as the gray curve, while the luminosity radiated at the surface $L_\mathrm{surf}$ is shown in red. The x-axis shows time elapsed, with $t=0$ corresponding to the birth of the original brown dwarf to which mass is later added. The plots begin at the start of the accretion phase. The end of the accretion phase is marked by a pair of circles on each line, and is either $10^7$ or $10^9$ years after the start of the accretion phase, or at $t=0$ in the case of the ordinary brown dwarf and star. Time periods during which an object has a frozen core or luminosity plateau (see text) are marked as blue and light blue lines respectively. Brown dwarf-like objects, namely the beige dwarfs in the first column and the brown dwarf, have surface luminosities that exceed the nuclear luminosity in the long run, while the pink and maroon dwarfs (right 3 columns) and the star arrive at an equilibrium where $L_\mathrm{surf}=L_\mathrm{nuc}$, often after multiple Gyr.}
\label{fig:gallery_Lt}
\end{figure*}

We recreate Figure \ref{fig:schematic} with our MESA models in Figure \ref{fig:massentropy} by plotting the models' mass and $\psi$ at the end of the accretion phase. Each point is then given a colour and symbol according to its final fate as follows:
\begin{itemize}
    \item If at the end of Run 3, i.e. long after the accretion phase, the object has a surface luminosity $L_\mathrm{surf}$ more than 0.1 dex greater than the nuclear luminosity $L_\mathrm{nuc}$, it is classified as essentially a {\bf brown dwarf}, or if its mass is above the HBL, a {\bf beige dwarf}. In either case, it is marked as a yellow circle in Figure \ref{fig:massentropy}.
    \item If the object is not a brown dwarf as defined above, i.e. it approaches $L_\mathrm{nuc} \approx L_\mathrm{surf}$ at the end of Run 3, but it becomes more degenerate over the course of Run 3, meaning that has moved leftwards in the Figure after its accretion phase, it is called a {\bf maroon dwarf} and marked with a maroon pentagon.
    \item If the object fits neither of the previous categories, i.e. it becomes less degenerate over time and reaches the main sequence, it is a {\bf pink dwarf} of some variety and will be marked in the Figure according to one of the following three possibilities.
    \item If the object has a {\bf luminosity plateau} of any duration, meaning that the surface luminosity changes slowly, $0 <d\log_{10} L/dt<0.3\ \mathrm{dex}\ \mathrm{Gyr}^{-1}$ at a luminosity within $10^{-4.9} L_\odot < L_\mathrm{surf} < 10^{-4.68} L_\odot$ (see the Appendix \ref{sec:appendix}) it is denoted with a diamond. Its colour is pink by default, but may be (and typically is) overridden by the next criterion.
    \item If the object has a {\bf frozen core}, meaning that the star is not fully convective, and this phase lasts for at least 100 Myr after the mass accretion phase, it is coloured according to the colourbar. The exact definition we use is that a star has a frozen core when its convective core mass as reported in the MESA history file as \texttt{mass\_conv\_core}, is less than $0.95$ times the star's total mass. If the object does not have a luminosity plateau, we use a square marker, but if it does, we use a diamond (see above).
    \item If a pink dwarf has neither a luminosity plateau nor a frozen core as defined above, it is simply shown as a pink star.
\end{itemize}

A clear dividing line between the beige/brown and pink dwarfs is visible, which is depicted in Figure \ref{fig:schematic} by the unstable left hand side of the parabolic curve.The pink dwarfs closest to the dividing line have frozen cores for a greater duration than those further away. Evidently the further in to the middle of the parabola one goes in the course of the accretion phase, the shorter the duration of the frozen core phase. Our cartoon version of the Figure appears to work remarkably well despite its simplifying assumption of constant entropy within the star.

A handful of the objects with luminosity plateaus do not appear to have frozen cores on this plot. In Figure \ref{fig:massentropy}, these points are diamonds but are simply pink and not on the colour bar. These models have been identified to have plateaus and do actually have a frozen core for a small amount of time. However, they have not met the frozen core condition for long enough to be classified as having a frozen core and are therefore not on the colour bar.

Figure \ref{fig:massentropy} also includes a set of isochrones derived from an ordinary set of MESA models with no mass accretion phase. By around 1 Gyr, stars above $\sim 0.085 M_\odot$ have settled onto the main sequence, while lower-mass stars continue becoming more degenerate with $L_\mathrm{surf}>L_\mathrm{nuc}$. Brown dwarfs of course become steadily more degenerate over time pulling the low-mass ends of the isochrones further left. We show three sets of isochrones, including modifying the metallicity to about 10 times solar (blue lines) and modifying the atmospheric boundary condition from \texttt{table} to \texttt{T\_tau}. Adjusting the metallicity has a much greater effect than the adjustment of the atmosphere model, but the isochrones are qualitatively similar in all cases. 

Several other features of Figure \ref{fig:massentropy} are noteworthy. First, it should be noted that while the x-axis $\psi$ has a one-to-one correspondence with the time that the original brown dwarf is allowed to cool in Run 1, the pattern of points from Figure \ref{fig:runs} is not quite the same as what is shown in Figure \ref{fig:massentropy}. There are some similarities, particularly at low values of $\psi$, but many of the ``Other pink dwarfs'' are clustered close to the main sequence. This is because when the accretion is slow, i.e. the mass is added over the course of $10^9$ years, a substantial amount of evolution occurs during the accretion phase, and the points in Figure \ref{fig:massentropy} are shown at the end of the accretion phase. In contrast, when the accretion time is only $10^7$ years, the models evolve essentially vertically during the accretion phase yielding a recognisable pattern from Figure \ref{fig:runs}. 

Next, one can read off from Figure \ref{fig:massentropy} that maroon dwarfs typically end their accretion phase rightward of the 1 Gyr isochrone, except in the lowest-mass cases. These models therefore end up in the same region as ordinary low mass pre-main sequence stars.

Having established that our schematic diagram corresponds well to the behaviour of the numerical models, we can now examine the detailed evolution of a number of individual models. We select 5 random examples each of beige dwarfs, pink dwarfs with a frozen core phase, pink dwarfs without a frozen core phase, and maroon dwarfs. In Figure \ref{fig:gallery_Lt} we show the time evolution of $L_\mathrm{nuc}$ and $L_\mathrm{surf}$. We also indicate the presence of frozen cores and plateaus, and the time at which accretion stops in these models. For comparison we also include an ordinary $0.065 M_\odot$ brown dwarf and an ordinary $0.085 M_\odot$ star.

All 20 objects in the ``gallery'' began as brown dwarfs, with $L_\mathrm{surf}>L_\mathrm{nuc}$. All 10 of the example pink dwarfs shown (central columns) end their accretion phase with the inverse true, namely $L_\mathrm{nuc}>L_\mathrm{surf}$ just as expected given their location inside the parabola. After some time period of order 1 Gyr, these objects settle into equilibrium with $L_\mathrm{nuc}=L_\mathrm{surf}$, reaching the main sequence. This corresponds roughly to the Kelvin-Helmholtz timescale of objects at the end of the main sequence. In this transient phase, the pink dwarf may either smoothly approach the new equilibrium luminosity of order $10^{-3} L_\odot$, or spend an additional period of time with its luminosity barely changing above $\sim 10^{-5} L_\odot$. These periods are typically marked by blue lines, indicating that the convective core of the object does not encompass 95\% of the mass of the object, what we ultimately believe to be the cause of these features (see below). Some of these cases additionally have a luminosity plateau (light blue lines), though not always, since often the surface luminosity does increase noticeably. Note that an ordinary $0.065\ M_\odot$ brown dwarf is also marked as having a frozen core, since after about 5 Gyr of evolution, its core becomes sufficiently degenerate that energy is no longer transported via convection. Similarly, beige dwarfs are characterized by a lack of convection (in their cores) that is unaffected by their gain in mass during the accretion phase, so they are also denoted as possessing frozen cores.

The beige dwarfs, despite now being above the hydrogen burning limit, and despite their nuclear luminosity increasing substantially during the accretion phase, end up with $L_\mathrm{nuc}<L_\mathrm{surf}$, with both luminosities steadily decreasing over multiple Gyr timescales. In one case, the third example, there is actually a period where $L_\mathrm{nuc}$ exceeds $L_\mathrm{surf}$ for a brief period. This model happens to be just below the boundary between beige and pink dwarfs, i.e. the left-hand branch of the $L_\mathrm{nuc}=L_\mathrm{surf}$ parabola, and shows that the schematic diagram and the associated story is not perfect. Having $L_\mathrm{nuc}>L_\mathrm{surf}$ for a brief period is not {\em guaranteed} to produce a pink dwarf, though that is the usual outcome.

In the third example down in the other pink dwarfs column, there looks to be a small time when it has a frozen core. It has not been called a frozen core pink dwarf due to the duration of the frozen core being less than 100 Myr thus not being long enough to meet our definition. Luminosity plateaus are only identified in periods during which an object has a frozen core.

\begin{figure*}
\centering 
\includegraphics[scale=0.8]{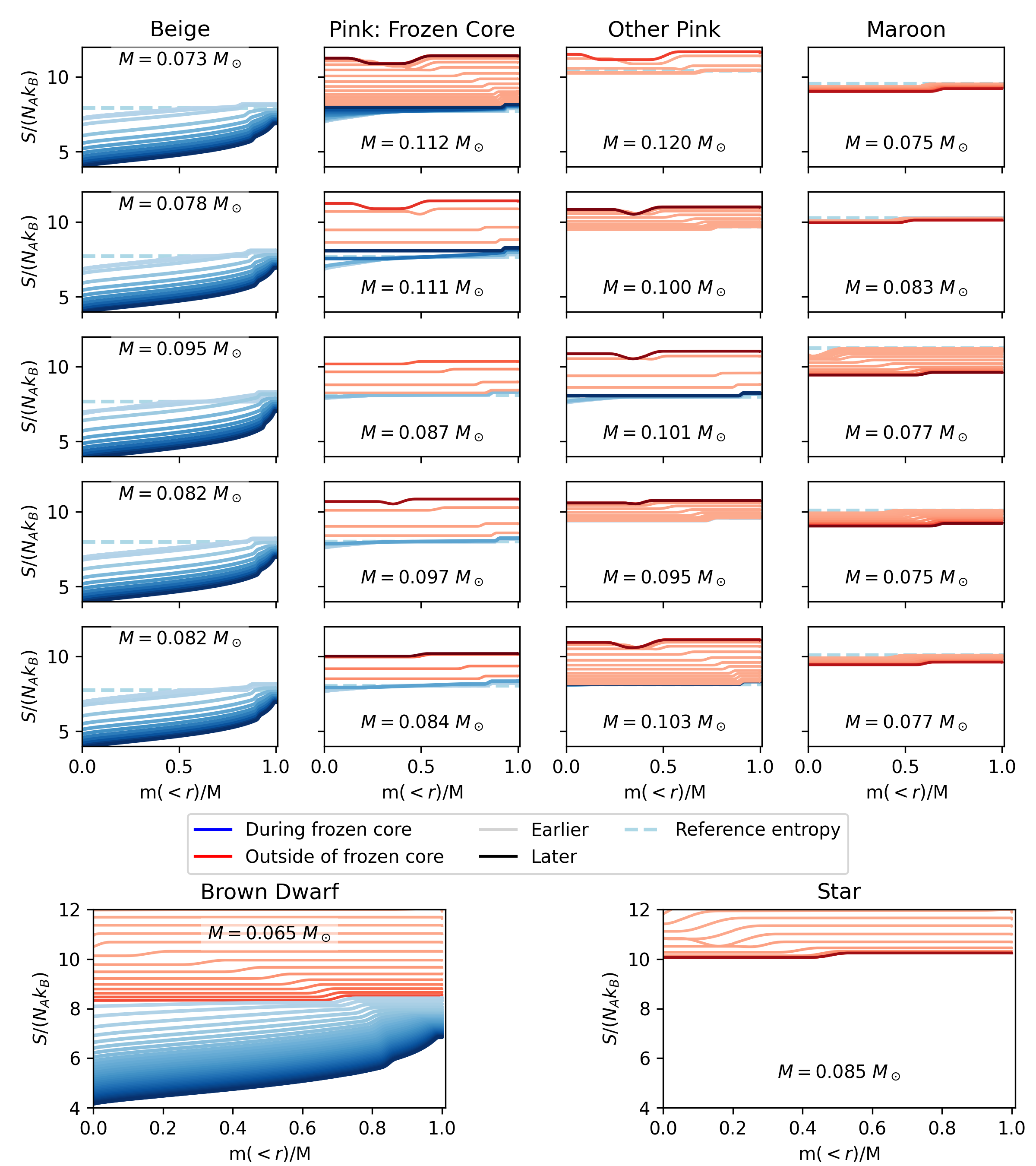}
\caption{The entropy profiles of the various example objects. The x-axis is the Lagrangian mass coordinate $m(<r)$ of the star normalised by the mass of the object $M$. The y-axis is the specific entropy divided by the product of Avogadro's number $N_A$ and Boltzmann's constant $k_B$, yielding a dimensionless number. Blue lines correspond to snapshots which are classified as having a frozen core, namely that the convective core mass is $<0.95\ M$. Red lines correspond to snapshots where this is not true, i.e. the object is fully convective. Within the red or blue lines, darker lines correspond to later times. Snapshots are only shown after the end of the accretion phase. These models are the same as those shown in Figure \ref{fig:gallery_Lt}.}
\label{fig:gallery_entropy}
\end{figure*}   

We have already alluded to the fact that the periods corresponding to slower-than-expected changes in $L_\mathrm{surf}$ have to do with a lack of convection in the core of the star. This seems to be a good hypothesis based on Figure \ref{fig:gallery_Lt}, where the end of this phase for pink dwarfs with frozen cores (second column) exactly corresponds to a dramatic upturn in the surface luminosity. To examine this process in more detail, we look at profiles of the same example objects shown in Figure \ref{fig:gallery_Lt}. In particular, in Figure \ref{fig:gallery_entropy} we plot the entropy as a function of the mass coordinate of the star, $m(<r)/M$. Flat entropy profiles are indicative of the action of convection, so we can immediately see that beige dwarfs, old brown dwarfs, and pink dwarfs with frozen cores have some period of time (shown as blue profiles) where the entropy is substantially lower in the core than at the surface. Near the main sequence, stars have nearly flat entropy profiles, with small-scale bumps corresponding to changes in the ionization state of the gas. This makes it clear why the analytic model from \cite{JohnLoeb} based on \citet{auddy2016} and represented schematically in Figure \ref{fig:schematic} does an excellent job describing the evolution of these objects, even away from the equilibrium curve (the parabola). The assumption of constant entropy directly links the surface and core conditions, even when the object is not in thermal equilibrium. For particularly old brown dwarfs, and beige dwarfs, with degenerate cores, the model may not always hold because of the existence of an entropy gradient. This explains why in rare cases (e.g. the third row in the beige dwarf column of Figure \ref{fig:gallery_Lt}) the models do not strictly follow the rule that $L_\mathrm{nuc}>L_\mathrm{surf}$ moves the object further rightward within the parabola. The momentarily larger nuclear luminosity is not enough to restore a flat entropy profile and therefore convective energy transport. 

To get a sense of the energy necessary to produce a flat entropy gradient, we show a dashed blue line at the entropy where the entropy gradient is flat at the end of the accretion phase, i.e. the entropy in the convective part of the star. In practice this location is defined as the smallest value of $m(<r)$ such that $dS/dm < 1.25 M_\odot^{-1}$ in the first snapshot immediately after the end of the accretion phase. We will return to this reference entropy below to estimate the timescale of the frozen core phase.

We saw a clear boundary between pink and beige dwarfs in Figure \ref{fig:massentropy}. Comparing the left two columns, we see that at the end of the accretion phase (the lightest blue of the profiles), the beige dwarfs tend to have a larger gap between the reference entropy and the actual entropy in their cores. In this sense, the dividing line between beige and pink dwarfs can be understood as part of the same phenomenon that gives rise to luminosity plateaus. In order for the object to increase its surface luminosity, it must be able to activate convection and ``unfreeze'' its core, but the energy required to do so is too great for the beige dwarfs, and takes some time for the pink dwarfs.

\begin{figure*}
\centering 
\includegraphics[scale=0.8]{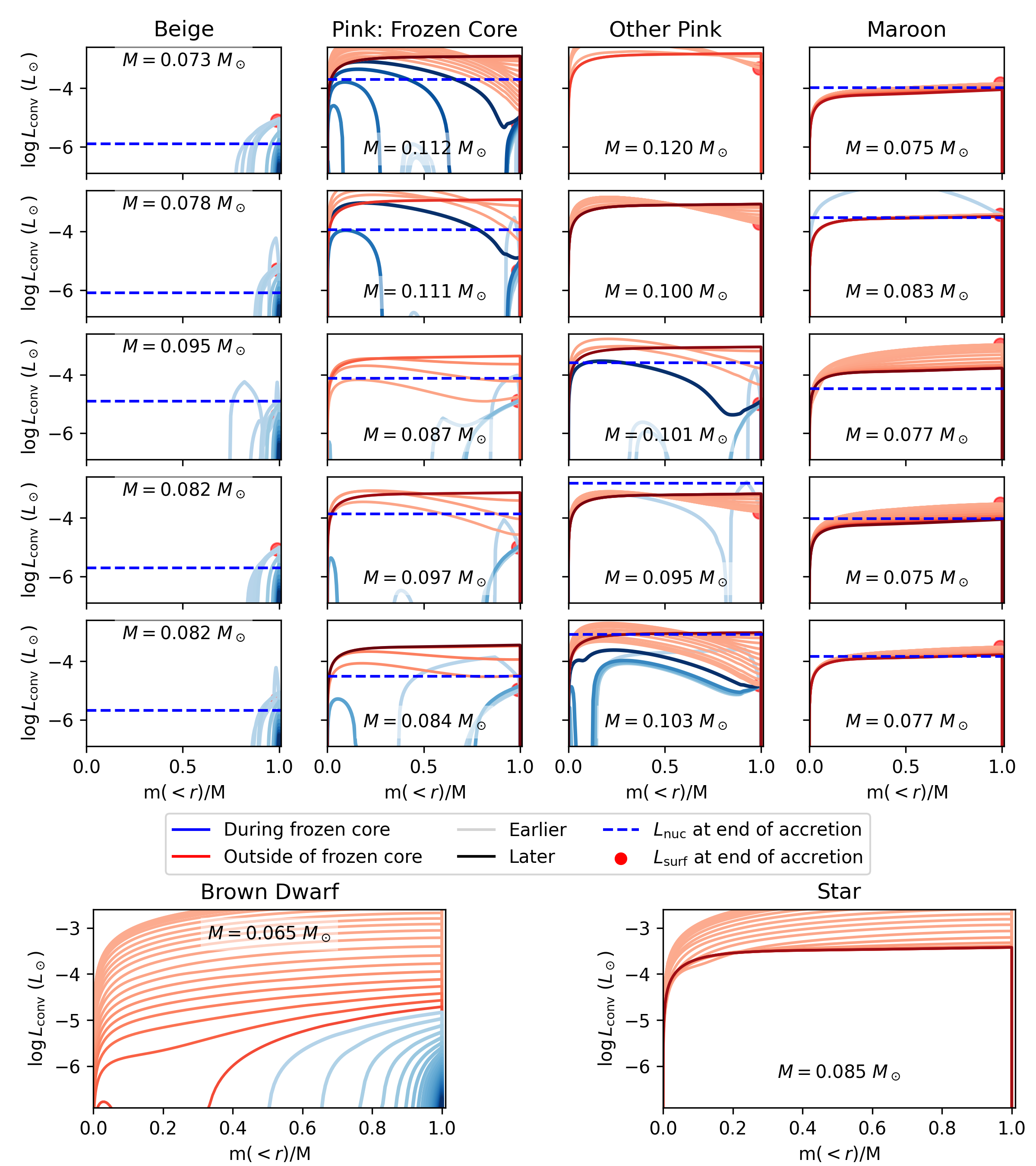}
\caption{Convective luminosity profiles. For the same objects shown in Figures \ref{fig:gallery_Lt} and \ref{fig:gallery_entropy}, we show their convective luminosities as a function of the Lagrangian mass coordinate of the object $m(<r)$ normalised by the mass of the object $M$. As with the entropy profiles, times at which the object is in a frozen core phase are coloured blue; when the objects are not in this phase, the profiles are red. Additionally we show for reference the nuclear luminosity at the end of accretion as a blue dashed line, and the surface luminosity at the same time as a red dot near the surface. Notably during the frozen core phase, there is typically a large gap where the star is not convective, despite being convective near the surface and sometimes near the core. Over time, in the pink dwarfs with frozen cores, the convective regions join together and the profiles flatten, meaning energy from the core is being transported to the surface in thermal equilibrium. Prior to this phase though, substantial energy is being deposited into the interior of the object (when $d L_\mathrm{conv}/dm < 0$).}
\label{fig:gallery_log_Lconv}
\end{figure*}  

After accretion stops, we see that all pink dwarfs, both with frozen cores and without, have their entropy increase overtime throughout the star. This behaviour differs from maroon dwarfs which instead have their entropy decrease over time, much like the regular star example in Figure \ref{fig:gallery_entropy}.

We can explicitly examine the energy flow in these objects by plotting profiles of the convective luminosity $L_\mathrm{conv}$, the luminosity passing through each shell of the star that is carried convectively. In an ordinary star, this quantity rises in the center as more of the region where nuclear reactions occur is contained within $m(<r)$, until the profile flattens and remains constant until it reaches close to the photosphere. In a brown dwarf, convection gradually shuts off in the core and the luminosity being transported to the surface is coming mostly from the outer layers of the star losing their thermal energy.

Pink dwarfs with frozen cores tend to begin convection in the core, but during the frozen core phase, the profiles eventually begin sloping downwards, meaning that the convective energy is being deposited at that location in the star. That is, the star is heating up and unfreezing from the inside out, until finally convection is able to transfer energy from the core all the way to the surface continuously. At this point the frozen core phase ends and the luminosity reaching the surface of the star rapidly climbs to of order $10^{-3} L_\odot$. The core had been producing these large luminosities earlier, but all of that energy was going in to heating the star and reactivating convection.

\section{Discussion}
\label{sec:discussion}

We have found that pink dwarfs can take a substantial amount of time to traverse the space in the $M-\psi$ diagram between the point at which they have stopped gaining mass and the main sequence. This is because it takes some time for their cores to heat up and ``unfreeze'', i.e. for convection to begin again in their cores. As a result, these objects may remain at comparatively low surface luminosities for extended periods of time. In Figure \ref{fig:massentropy} we saw that in our MESA models, the amount of time it took the object to unfreeze its core was closely related to its distance from the $L_\mathrm{nuc}=L_\mathrm{surf}$ curve -- the closer the object was to this curve at the end of its mass accretion phase, the longer its core remained frozen.

\begin{figure}[]
\centering 
\includegraphics[scale=0.6]{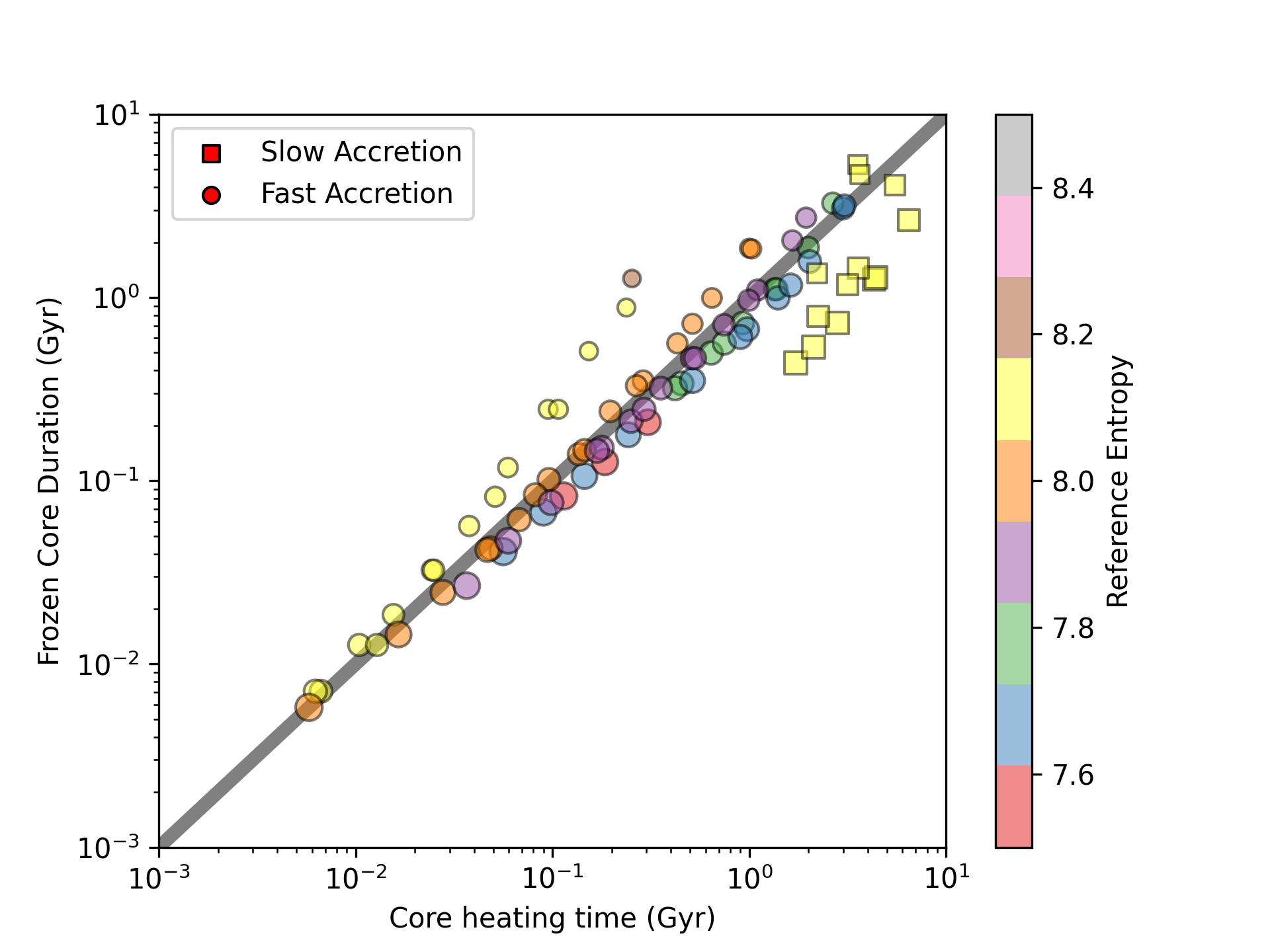}
\caption{The duration of the frozen core of an object against the core heating time for that object. The core heating time is a timescale defined by Equation \ref{eq:coreheating}. The slow accretion points have had mass accreted over $10^9$ years and the fast accretion time is $10^7$ years. The points are coloured with respect to their reference entropy, which is the entropy where the inner most point of the object has a flat entropy profile when accretion stops (see also Figure \ref{fig:gallery_entropy}). The size of the points is dependent on the mass of the object at the end of the model, the larger the point, the more massive the object. The points lie close to $y=x$ (the gray line).}
\label{fig:khsmore}
\end{figure}

We now attempt to quantify the timescale that the object retains its frozen core. The frozen core timescale, as discussed in the context of Figures \ref{fig:massentropy} and \ref{fig:gallery_Lt} is the time period during which the convective core mass is less than 95\% of the star's mass, starting from the end of the accretion phase. We expect this to be closely related to the timescale required to heat the core, which we estimate as the deficit in thermal energy relative to a fully convective state divided by the rate of energy injection by nuclear reactions,
\begin{equation}
\label{eq:coreheating}
    t_\mathrm{core\ heating} = \frac{1}{L_\mathrm{nuc}}\int_0^{m_\mathrm{ref}} (S - S_\mathrm{ref}) C_V T dm
\end{equation}
Each quantity in this equation is evaluated at the end of the accretion phase, and shown relative to the measured frozen core timescale in Figure \ref{fig:khsmore} for all models with a frozen core phase. The points lie close to the 1:1 line, indicating that this is indeed a meaningful description of what is happening in the cores of these objects. The cores remain frozen until the nuclear luminosity is able to supply enough energy to heat the object to a constant-entropy state, at which point convection can operate, and the nuclear and surface luminosities are connected. Once this happens, the two luminosities converge on a thermal timescale (see Figure \ref{fig:gallery_Lt}).

\begin{figure}[t]
\includegraphics[scale=0.45]{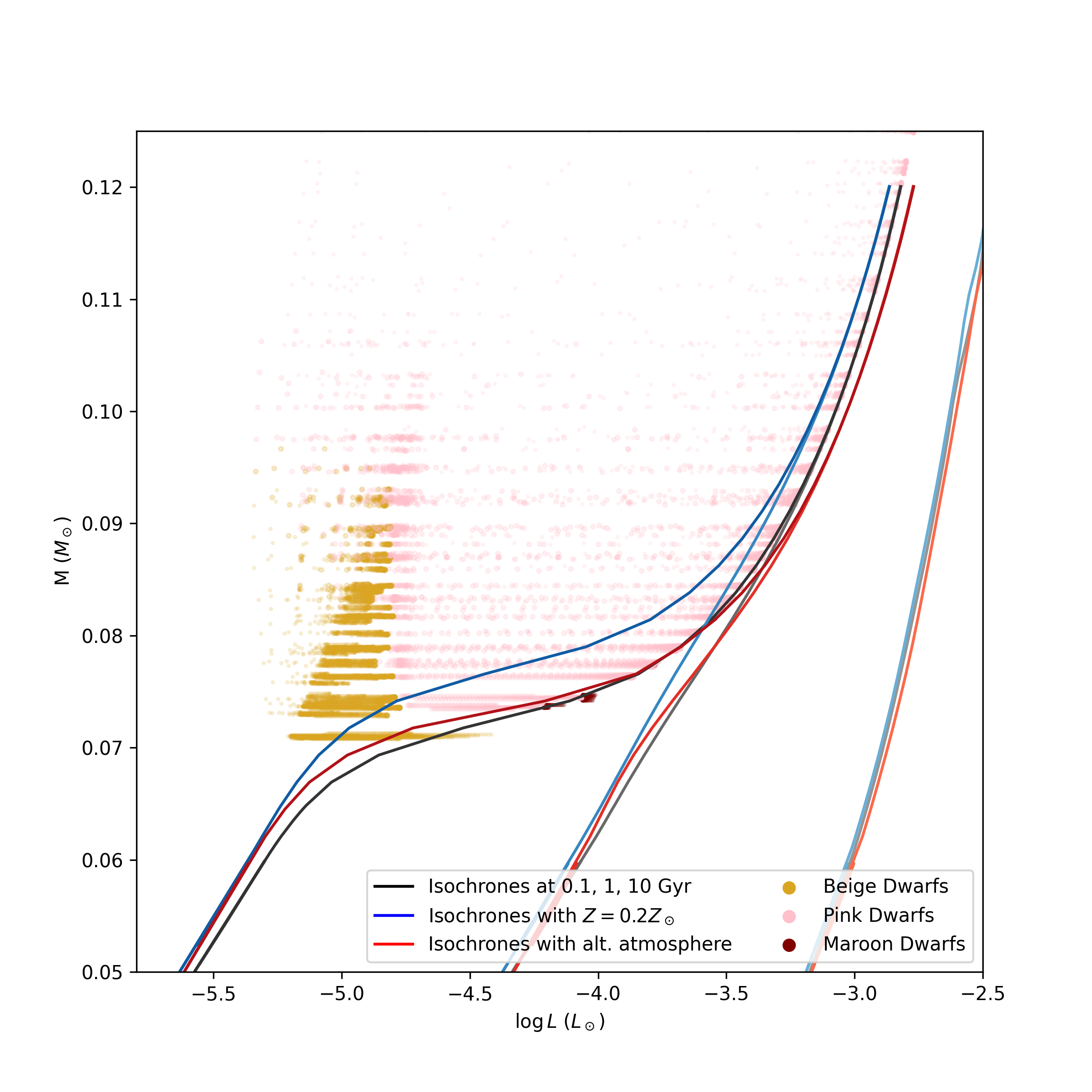}
\caption{
Mass and luminosity of beige, pink, and maroon dwarfs after accretion has stopped. For each object, the luminosity has been plotted every 176 Myr between 5 Gyr and 13.8 Gyr (corresponding to 50 points per object if the end their accretion phase earlier than 5 Gyr). There is a region of this plot that has a higher density of pink and beige dwarfs due to their lower rate of change of luminosity, seen in Figure \ref{fig:luminosity_dL}. 
Three versions of isochrones at 0.1, 1, and 10 Gyrs have been plotted, with the 10 Gyr isochrones showing the location of the main sequence. Black is standard with solar metallicity and tau tables 100. Blue has an adjusted metallicity with $Z = 0.2Z_\odot$ and red has had \texttt{atm\_option} changed to \texttt{T\_tau} from \texttt{table}.}
\label{fig:log_L_star_mass}
\end{figure}

This extended period during which the luminosity of some pink dwarfs remains far smaller than would be expected for an object at the same mass on the main sequence presents an observational opportunity to identify pink dwarfs, and the closely-related beige dwarfs, which remain at low luminosities indefinitely. Beige and pink dwarfs occupy a distinct region in a diagram of mass vs. luminosity (Figure \ref{fig:log_L_star_mass}, which shows pink, beige, and maroon dwarfs in this space compared to the 0.1, 1, and 10 Gyr isochrones of sets of ordinary MESA models where no mass is added to the objects). Around luminosities of $10^{-5} L_\odot$ is a concentration of beige and pink dwarfs, in a region inaccessible by single-star evolution, except at substantially subsolar abundances. The main difficulty with using this diagram observationally is that the exact location of the isochrone should not be trusted, as its exact position depends on fine details of the modelling, which similarly affect the position of the hydrogen burning limit, visible for our simulations in the 10 Gyr isochrones as the region where luminosity begins to fall only gradually with mass around $0.07\ M_\odot$. The Solar abundance \citep{dieterich2014} and the equation of state \citep{2023Chabrier}, for instance, each affect the inferred location of the HBL.

Nonetheless, it may be possible to estimate luminosities and masses of objects with observational data to search for those that lie above and to the left of these isochrones. Estimating each of these quantities is not without its challenges. The mass requires a binary companion, which in turn may obscure the brightness of the brown dwarf and make estimating its luminosity challenging. We expect that the primary route by which pink and beige dwarfs form is via mass transfer from a star about the mass of the Sun \citep{DorsaJohn}. This allows the brown dwarf enough time to cool down before the evolution of the primary star instigates the mass transfer. \citet{DorsaJohn} found that mass transfer via wind-assisted Roche lobe overflow \citep{mohamed2007} was sufficient to push massive brown dwarfs above the hydrogen burning limit. Mass transfer within a common envelope phase when the donor star is on either the RGB or AGB branch may also provide sufficient mass transfer to leave the brown dwarf as a beige or pink dwarf \citep{livio1984,maxted2006}.

In either case, the beige or pink dwarf will be left in orbit around a typical $\sim 0.56\ M_\odot$ \citep{cummings2018} white dwarf, so the search for beige or pink dwarfs is likely to be a matter of finding faint low-mass companions of white dwarfs, and estimating the mass and luminosity of the companion. Finding candidate low-mass companions to white dwarfs is possible in photometric surveys owing to the large difference in temperature between the white dwarf and the low-mass object \citep[e.g.][]{steele2011}, although the infrared excesses uncovered in these searches can also be due to debris disks \citep[e.g.][]{farihi2012}. The mass of the companion may be estimated with radial velocities \citep[e.g.][]{koester2001A&A,crumpler2025}, though only up to a factor of the sine of the inclination, and with some difficulty owing to the broadness and asymmetry of lines in white dwarf spectra \citep[e.g.][]{napiwotzki2020, arseneau2024}. The mass may also be constrained astrometrically with {\em Gaia}, which does contain a number of white dwarfs in the non single star catalogue. Meanwhile binary SED fitting \citep[e.g.][]{prvsa2005,bayo2008,thompson2021,jadhav2025} may be used to estimate the luminosities, though again the infrared portion of the spectrum may contain contributions from debris disks, and the absolute luminosity is subject to distance uncertainties, which can be substantial.

Beyond mass and luminosity, it is also possible to find chemical signatures of mass transfer from AGB stars to their companions, namely barium \citep[e.g.][]{boffin1988,han1995,krynski2025} and carbon stars \citep[e.g.][]{ardern2025}. However, carbon enhancements are typically only detectable at very low metallicities where the initial carbon abundance of the recipient star is low enough that an enhancement is detectable \citep[e.g.][]{yamaguchi2025}. Meanwhile barium enhancements have not been detected in stars much smaller than $1\ M_\odot$ \citep[e.g.][]{kong2018}, though this is likely due to the practical difficulties of estimating barium abundances in lower-mass stars. The lowest-mass barium star in \citet{yamaguchi2025} is $\approx 0.6\ M_\odot$, far more massive than where we expect that either pink or beige dwarfs would be able to maintain luminosities of order $10^{-5} L_\odot$. We therefore anticipate that {\em Gaia}, via its astrometric characterization of the companion star and the constraints it places on the distance to these systems is the most promising avenue of searching for pink or beige dwarfs, particularly with radial velocity follow-up \citep[e.g.][]{rogers2024}. Chemical signatures may still be detectable though, since the convective layers near the surface of both beige dwarfs and pink dwarfs with frozen cores often include only $\sim10\%$ of the object's mass (see Figure \ref{fig:gallery_log_Lconv}).

\section{Conclusion}
\label{sec:conclusion}

Stellar objects near the hydrogen burning limit have traditionally been split into red and brown dwarfs, with the dividing line being the HBL itself. When enough mass is added to a brown dwarf to push it over the HBL, we have divided the possibilities into the following three categories (see Figure \ref{fig:schematic}) based on the timing and quantity of the mass added:
\begin{itemize}
    \item {\bf Beige Dwarfs} -- named for their similarity to both white dwarfs and brown dwarfs. These objects never reach the $L_\mathrm{nuc}=L_\mathrm{surf}$ line, so once they have gained their mass they simply cool and become more degenerate over time. They are essentially brown dwarfs but more massive.
    \item {\bf Maroon Dwarfs} -- named for their similarity to red dwarfs. These objects gain their mass early in their lives before they have had a chance to cool past the minimum of the $L_\mathrm{nuc}=L_\mathrm{surf}$ curve. They therefore immediately become indistinguishable, except perhaps chemically, from an ordinary pre-main sequence low-mass star.
    \item {\bf Pink Dwarfs} -- named for their similarity to beige dwarfs, having passed through the same region of $M-\psi$ parameter space, and to red dwarfs. These objects pass through the $L_\mathrm{nuc}=L_\mathrm{surf}$ curve as they are gaining mass, meaning that after they have stopped gaining mass, their cores will heat up, eventually bringing the object to the main sequence.
\end{itemize}

When material is added to a brown dwarf, its fate therefore depends not just on the final mass of the object, but on how degenerate its core is at the time the mass is added. The pink dwarfs, even though they will eventually become main sequence stars, may spend of order a Gyr with substantially lower luminosities than stars of their new mass. Just like beige dwarfs, they occupy a region above and to the left of a standard isochrone in mass vs luminosity space (Figure \ref{fig:log_L_star_mass}), which is in principle observable, particularly because these objects most likely exist in binaries. We present several promising candidates in forthcoming companion work (Rusk et al, in prep). Successful identification of candidate beige or pink dwarfs would be a remarkable confirmation of this theoretical work, and may allow nontrivial constraints on the physics of mass transfer and common envelope evolution.

\begin{acknowledgements}
      We are grateful for helpful conversations with Adam Burgasser, Jackie Faherty, Kelle Cruz, Louise Nielsen, Karen Pollard, Federico Marocco, Anke Ardern-Arentsen, and Yixiao Zhou. 
\end{acknowledgements}

\appendix
\section{Luminosity Plateau Definition}\label{sec:appendix}
\begin{figure}[t]
\centering 
\includegraphics[scale=0.45]{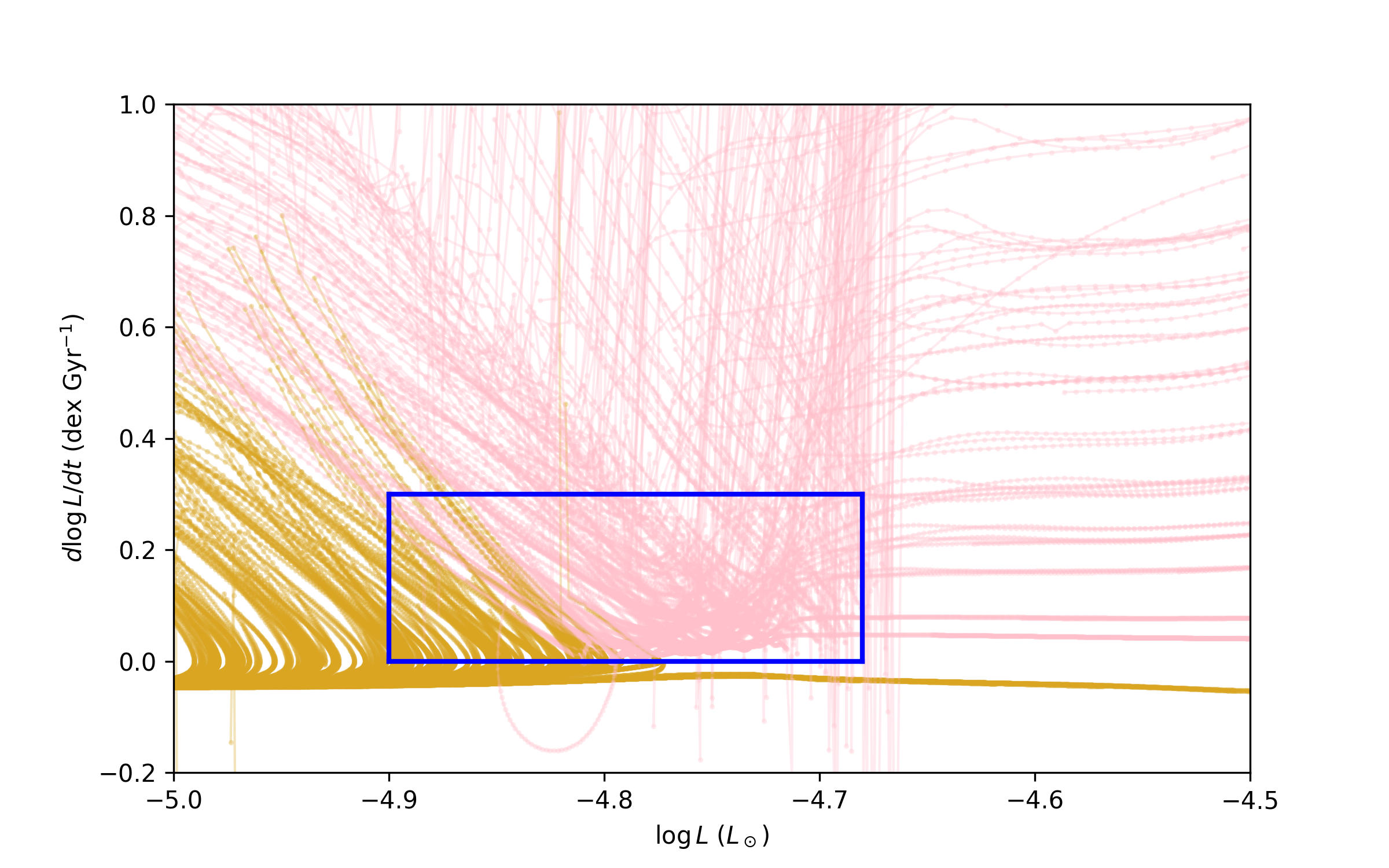}
\caption{The rate of change of the surface luminosity of a pink (shown in pink) or beige (shown in goldenrod) dwarf between 5 and 13.8 billion years, with points spaced 11 million years apart, depending on its surface luminosity. When $d\mathrm{log}L/dt$ of a pink or beige dwarf is near zero, their surface luminosity has plateaued. We have selected a region of this plot, shown by the blue box, where we defined an object to have a plateau if/when it passes through that region. Figure \ref{fig:gallery_Lt} has examples of beige and pink dwarfs' luminosity over time and some of them have a plateau and would have passed through the blue box and some of them do not.}
\label{fig:luminosity_dL}
\end{figure}

Our definition of a luminosity plateau is dependent on the rate of change of surface luminosity. We plot the trajectory of beige and pink dwarfs in a space of $d\log L_\mathrm{surf}/dt$ against $L_\mathrm{surf}$ in Figure \ref{fig:luminosity_dL}. There is a clear feature where the pink dwarfs all approach their minimum rate of change in luminosity around the same luminosity value of $\sim 10^{-4.75}L_\odot$. As each one continues to gain luminosity, this may happen slowly or quickly. To pick out this local minimum in the rate of change of the luminosity, we define models in the blue rectangle to be experiencing their plateau phase.

\bibliography{references}
\bibliographystyle{aasjournal}

\end{document}